\documentclass[prc,aps,nofootinbib,showpacs,preprintnumbers,amsmath,amssymb,floatfix]{revtex4}

\usepackage{times}
\usepackage{hyperref}

\usepackage{graphicx}
\usepackage{dcolumn}
\usepackage{bm}
\usepackage{slashed}
\usepackage[a0paper]{geometry}
\usepackage{anysize}
\marginsize{2.5cm}{2cm}{2.5cm}{2.5 cm}
\usepackage{amsmath}
\usepackage{amsfonts}
\usepackage{amssymb}
\usepackage{slashed}
\usepackage{mathrsfs}
\usepackage{graphicx}
\usepackage{epsfig}
\usepackage{enumerate}
\usepackage{amsmath}
\usepackage{pdfpages}
\usepackage{setspace}
\usepackage[utf8]{inputenc}
\usepackage{bm}
\usepackage{float}
\usepackage{epstopdf}

\newcommand{\pr}[1]{ \mathcal{P}^{(#1)}}

\begin{document}

\title{Contribution of $\bm{SU(3)}$ quadratic Casimir squared to rotational bands in the interacting boson model}

\author{
V{\'\i}ctor Miguel Banda Guzm\'an
}
\affiliation{
Instituto de F{\'i}sica y Matem\'aticas, Universidad Michoacana de San Nicol\'as de Hidalgo, Edificio C-3, Apdo.\ Postal 2-82, Morelia, Michoac\'an, 58040, M\'exico
}

\author{
Rub\'en Flores-Mendieta
}
\affiliation{
Instituto de F{\'\i}sica, Universidad Aut\'onoma de San Luis Potos{\'\i}, \'Alvaro Obreg\'on 64, Zona Centro, San Luis Potos{\'\i}, S.L.P.\ 78000, Mexico
}

\author{
Johann Hern\'andez
}
\affiliation{
Instituto de F{\'\i}sica, Universidad Aut\'onoma de San Luis Potos{\'\i}, \'Alvaro Obreg\'on 64, Zona Centro, San Luis Potos{\'\i}, S.L.P.\ 78000, Mexico
}

\date{\today}

\begin{abstract}
Rotational bands are commonly used in the analysis of the spectra of atomic nuclei. The early version of the interacting boson model of Arima and Iachello has been foundational to the description of rotations in nuclei. The model is based on a unitary spectrum generating algebra $U(6)$ and an orthogonal (angular momentum) symmetry algebra $SO(3)$. A solvable limit of the model contains $SU(3)$ in its dynamical symmetry chain. The corresponding Hamiltonian is written as a linear combination of linear and quadratic Casimir invariants of all the algebras in the chain. Prompted by these facts, a Hamiltonian containing the $SU(3)$ quadratic Casimir squared is proposed to evaluate its effects on rotational bands. The additional term yields three undetermined parameters into the theory, which need be obtained from experiment. The lack of data does not allow one to perform a detailed numerical analysis, but a rather restricted one in terms of a one-parameter fit. Nevertheless, these additional terms provide a good description of rotational bands of two nuclei of interest, $^{156}\mathrm{Ga}$ and $^{234}\mathrm{U}$.
\end{abstract}

\pacs{}

\maketitle

\section{Introduction}

The atomic nucleus is a non-trivial many-body quantum system which has collective properties resulting in various deformed shapes. A strongly deformed nucleus rotates, exhibiting characteristic rotational band structures with noticeable regularity. A number of different methods have been used to study many-body quantum systems. One of the oldest methods is the shell model which assumes that a single nucleon behaves under the influence of a nuclear mean field. Formally, in this description nucleons populate shells that represent energy levels, going from lower to higher energies. When the number of nucleons equals 2, 8, 20, 28, 50, 82, and 126 the shells are completely filled. These so-called magic numbers represent nuclei with great stability.

A second method which has been widely used in the understanding of the collective behavior of nuclei is the interacting boson model (IBM) originally introduced by Arima and Iachello \cite{iach75,iach1,iach2,iach3,iach4,ibm}. The earliest version of the model, applied to even-even nuclei, describes the collective properties in terms of pairs of valence nucleons, which conform a set of interacting $s$ and $d$ bosons with positive parity and with angular momenta $L=0$ and $L=2$, respectively. The simplest version of the model, usually referred to as IBM-1, treats both types of nucleons the same so it assumes that there is only a single kind of boson, unlike IBM-2, which treats protons and neutrons separately. 

From its very construction, the IBM possesses a symmetry-based formulation and in particular, dynamical symmetries are essential to it. 
Dynamical symmetries arise when the Hamiltonian of a system can be written in terms of the Casimir (or invariant) operators of a chain of groups (or algebras). A remarkable application of the group (or algebra) chains is in the construction of bases in which the Hamiltonian can be diagonalized, or equivalently, in the construction of bases that transform as representations of the appropriate groups, labeling those states with the corresponding quantum numbers \cite{iach75}.

In the IBM-1, with $s$ and $d$ bosons, the spectrum generating algebra is $U(6)$ and it has $SU(3)$ as a subalgebra generating rotational spectrum \cite{iach75,iach2}. Rotational bands lie within irreducible representations (irreps) of $SU(3)$, which are subspaces invariant under the generators of the group. The IBM-1 Hamiltonian in thus written as a linear combination of linear and quadratic Casimir invariants of all the algebras in a given chain \cite{iach75}. Those terms are usually refer to as $1$- and $2$-body terms, respectively. In most calculations only up to two-body terms are retained.

The aim of the present paper is to introduce a $4$-body contribution to the IBM-1 Hamiltonian to study its effects on the rotational bands of nuclei. This contribution will thus be proportional to the $SU(3)$ quadratic Casimir squared.

This paper is organized as follows. In Sec.~\ref{sec:overview} some necessary material on the IBM is reviewed in order to introduce notation and conventions. This includes elementary definitions about dynamical symmetries. In Sec.~\ref{sec:2body} the Hamiltonian for 2-body structures is reviewed, to immediately discuss the $4$-body Hamiltonian in Sec.~\ref{sec:4body}. In Sec.~\ref{sec:fit} a fit to experimental data for rotational bands is presented, via a one-parameter fit. Results and conclusions are given in Sec.~\ref{sec:concluding}. The paper is complemented by two appendices, where some useful material can be found.

\section{\label{sec:overview}A survey on the IBM}

In this section, an elementary survey on the IBM is provided in order to introduce notation and conventions. For this purpose, the first concept to be reviewed is that of a dynamical symmetry. A comprehensive introduction on the subject can be found in Ref.~\cite{leviatan}, so only a few salient facts will be repeated here.

Dynamical symmetries arise when the Hamiltonian $H$ of a system can be written as a sum of commuting operators in the form
\begin{equation}
H = \sum_G a_G C_G,
\end{equation}
where $C_G$ are the Casimir operators of a chain of nested algebras\footnote{For ease of notation, no distinction between a group, for instance $U(6)$, and its algebra, $\mathfrak{u}(6)$, will be made, but denote both by a capital letter, $U(6)$.}
\begin{equation}
G_\mathrm{dyn} \supset G_1 \supset G_2 \supset \ldots \supset G_\mathrm{sym}. \label{eq:chaindyn}
\end{equation}
If this is the case, the spectrum can be solved in an explicit analytic form: The eigenstates $|\lambda_\mathrm{dyn},\lambda_1, \lambda_2, \ldots, \lambda_\mathrm{sym}\rangle$ and eigenvalues $E(\lambda_\mathrm{dyn}, \lambda_1, \lambda_2, \ldots, \lambda_\mathrm{sym})$ are labeled by quantum numbers $\lambda_\mathrm{dyn},\lambda_1,\lambda_2,\ldots,\lambda_\mathrm{sym}$, which characterize irreps of the algebras in the chain. Note that the condition of the nesting of the algebras in (\ref{eq:chaindyn}) is indispensable for constructing a set of commuting operators and hence for obtaining an analytic solution. For definiteness, $G_\mathrm{dyn}$ stands for the spectrum generating algebra of the system so operators corresponding to physical observables can be written in terms of its generators and $G_\mathrm{sym}$ is the symmetry algebra \cite{iach95}. Note also that the symmetry $G_\mathrm{dyn}$ is broken and the only remaining symmetry is $G_\mathrm{sym}$ which is the true symmetry of the problem. A $G_\mathrm{dyn}$ algebra can have several chains so analytic solutions can only be obtained whenever the Hamiltonian can be written in terms only of the Casimir operators of a given chain.

As for the IBM-1, it is based on a unitary spectrum generating algebra $G_\mathrm{dyn}=U(6)$ and an orthogonal (angular-momentum) symmetry algebra $G_\mathrm{sym}=SO(3)$. The Hamiltonian is expanded in the elements of $U(6)$ and consists of Hermitian, rotational-scalar interactions which conserve the total number of $s$ and $d$ bosons, $\hat{N}=\hat{n}_s+\hat{n}_d = s^\dagger s + \sum_m d_m^\dagger d_m$ \cite{leviatan}. The IBM-1 admits three chain algebras, namely,
\begin{equation}
U(6) \supset \left\{ \begin{array} {c} U(5) \supset SO(5) \\ SU_\pm(3) \\ SO_\pm(6) \supset SO(5) \end{array} \right\} \supset SO(3), \label{eq:fullchain}
\end{equation} 
which are known as the vibrational $U(5)$ \cite{iach1}, the rotational $SU(3)$ \cite{iach2}, and the $\gamma$-unstable $SO(6)$ limits \cite{iach4}. The $\pm$ labels attached to $SU(3)$ and $SO(6)$ in chains (\ref{eq:fullchain}) will serve as a reminder that those algebras have two different realizations depending on the phase choices for the $s$ and $d$ bosons \cite{shi}.

The present analysis is concerned with the study of rotational bands, so the chain algebra under consideration is
\begin{equation}
U(6) \supset SU_\pm(3) \supset SO(3), \label{eq:chain}
\end{equation}
where the algebras $SU_+(3)$ and $SU_-(3)$ correspond to prolate and oblate shapes, respectively

A Hamiltonian exhibiting the chain algebra (\ref{eq:chain}) can be constructed as
\begin{equation}
H^{SU(3)} = c_1 + c_2 C^{(1)}_{U(6)} + c_3 C^{(2)}_{U(6)} + c_4 C^{(2)}_{SU_\pm(3)} - c_5 C^{(2)}_{SO(3)}, \label{eq:h2su3}
\end{equation}
where $C^{(1)}_{U(6)}$ and $C^{(2)}_{U(6)}$ are the linear and quadratic Casimir operators of $U(6)$ and $C^{(2)}_{SU_\pm(3)}$ and $C^{(2)}_{SO(3)}$ are the quadratic Casimir operators of $SU_\pm(3)$ and $SO(3)$, respectively. Since a quadratic Casimir operator is obtained from the product of two creation and two annihilation operators, it is usually referred to as a $2$-body operator. In general, $n$-body operators are constructed from the product of $n$ creation and $n$ annihilation operators.

The eigenstates of the Hamiltonian (\ref{eq:h2su3}) can be represented by
\begin{equation}
|N,(\lambda_\pm,\mu_\pm),K_\pm,L\rangle,
\end{equation}
where $N$, $(\lambda_+,\mu_+)$, $(\lambda_-,\mu_-)$, and $L$ label the relevant irreps of $U(6)$, $SU_+(3)$, $SU_-(3)$, and $SO(3)$, respectively, and $K_+$ and $K_-$ are multiplicity labels \cite{leviatan}. The quantum numbers for $SU_-(3)$ and $SU_+(3)$ are not identical but are obtained from each other under the interchange $\lambda \leftrightarrow \mu$. Hereafter, only the $SU_-(3)$ labels are involved, so they will be loosely denoted by $(\lambda,\mu)$ for simplicity. No confusion should arise because irreps with $\lambda > \mu$ correspond to prolate shapes whereas irreps with $\lambda < \mu$ correspond to oblate shapes \cite{bon}.

Therefore, for a given $N$, there are several $SU(3)$ irreps with $(\lambda, \mu)$ defined as \cite{Pfeifer}
\begin{subequations}
\begin{eqnarray}
\mu & = & 0, 2, 4, \dots \\
\lambda & = & 2N-6l-2\mu, \qquad l = 0, 1, \dots, N,
\end{eqnarray}
\end{subequations}
and for a given $SU(3)$ irrep there are eigenstates with different $L$ expressed in terms of Elliott's quantum number $K$ \cite{elliott1,elliott2}
\begin{subequations}
\begin{eqnarray}
K & = & 0, 2, 4, \dots, \text{min}(\lambda,\mu), \\
L & = & \left\{
\begin{array}{lll}
0, 2, 4, \dots, \text{max}(\lambda,\mu), &\quad & \mbox{for} \quad K = 0, \\[3mm]
K, K+1, K+2, \dots, K+\text{max}(\lambda,\mu), & \quad & \mbox{for} \quad K > 0.
\end{array}
\right.
\end{eqnarray}
\end{subequations}

Since
\begin{equation}
C^{(2)}_{SO(3)} = \mathbf{L}^2,
\end{equation}
where $\mathbf{L}\equiv (L_1,L_2,L_3)$ is the angular momentum operator for a fixed value of $(\lambda,\mu)$, then for a given $SU(3)$ irrep there is a rotational band with energy levels spaced according to the eigenvalues of $\mathbf{L}^2$, that is to say $L(L+1)$. In the $2$-body Hamiltonian (\ref{eq:h2su3}), the energy difference between these rotational bands is given by the $SU(3)$ quadratic Casimir $C^{(2)}_{SU(3)}$. In general, the differences would be controlled by higher-order $n$-body $SU(3)$ invariants in the IBM Hamiltonian. Therefore, in order to obtain a more accurate phenomenological description of the nuclear spectra and be able to understand the role that higher-order terms in the chain group (\ref{eq:chain}) play, the aim of the present work is to propose a general $4$-body $SU(3)$ Hamiltonian. For this task, the $SU(3)$ projection operators introduced in Ref.~\cite{banda} come in handy.

\section{\label{sec:2body}IBM-1 in the rotational limit for $2$-body operators}

The building blocks of the IBM-1 are the creation and annihilation operators of the $s$ and $d$ bosons. Creation operators will be denoted as
\begin{equation}
s^\dagger, \quad d_m^\dagger, \qquad (m = 0, \pm 1, \pm 2) \nonumber
\end{equation}
so that the commutator of the creation and annihilation operators that are associated with the same boson state equals one, while all other commutators vanish.

Defining $b^\dagger_{l,m}$, with $l=0,2$ and $-l \leq m \leq l$, as \cite{ibm}
\begin{equation}
b^\dagger_{0,0} = s^\dagger, \quad b^\dagger_{2,2} = d^\dagger_{2}, \quad b^\dagger_{2,1} = d^\dagger_{1}, \quad b^\dagger_{2,0} = d^\dagger_{0}, \quad b^\dagger_{2,-1} = d_{-1}, \quad b^\dagger_{2,-2} = d^\dagger_{-2},
\end{equation}
the canonical commutation relations are now written as
\begin{equation}
[b_{l,m}, b^\dagger_{l^\prime,m^\prime} ] = \delta_{ll^\prime} \delta_{mm^\prime}, \quad [b^\dagger_{l,m}, b^\dagger_{l^\prime,m^\prime} ] = [b_{l,m}, b_{l^\prime,m^\prime} ]=0.
\end{equation}

In the IBM-1, a Hamiltonian that conserves the total number of bosons is constructed from $n$-body operators, which represent the interactions between bosons. Up to $2$-body operators, the most general Hamiltonian reads
\begin{equation}
H = E_0 + \sum_{\alpha \beta} c_{\alpha \beta} b_\alpha^\dagger b_\beta + \sum_{\alpha \beta \gamma \delta} c_{\alpha\beta\gamma\delta} b_\alpha^\dagger b_\beta^\dagger b_\gamma b_\delta, \label{eq:GH}
\end{equation}
where the Greek indices are a short-hand notation for the two indices in the creation and annihilation operators; for example, $\alpha = (l,m)$, and so on.

The eigenvalue problem for $H$ can be solved analytically for some particular cases. To illustrate this point, consider the algebra of the bilinear operators \cite{ibm}
\begin{equation}
G_{\kappa}^{(k)}(l,l^\prime) = [ b_l^\dagger \times \tilde{b}_{l^\prime} ]_{\kappa}^{(k)}, \label{eq:bilop}
\end{equation}
where $l,l^\prime = 0, 2$, $\tilde{b}_{l,m}=(-1)^{l+m}b_{l,-m}$, and the right-hand side of Eq.~(\ref{eq:bilop}) can be written as
\begin{equation}
[ b_l^\dagger \times \tilde{b}_{l^\prime} ]_{\kappa}^{(k)} = \sum_{m m^\prime}(lm l^\prime m^\prime|k\kappa) B_{lm;l^\prime m^\prime},
\end{equation}
where $(lml^\prime m^\prime |k\kappa)$ denote Clebsch-Gordan coefficients and $B_{lm;l^\prime m^\prime} = b^\dagger_{lm} \tilde{b}_{l^\prime m^\prime}$, whose commutation relations are obtained as
\begin{equation}
[ B_{lm;l^\prime m^\prime}, B_{\tilde{l}\tilde{m};\tilde{l}^\prime \tilde{m}^\prime} ] = (-1)^{l^\prime +m^\prime} \delta_{l^\prime \tilde{l}} \delta_{-m^\prime \tilde{m}} B_{lm;\tilde{l}^\prime \tilde{m}^\prime} - (-1)^{\tilde{l}^\prime +\tilde{m}^\prime} \delta_{l \tilde{l}^\prime} \delta_{m -\tilde{m}^\prime} B_{\tilde{l}\tilde{m};l^\prime m^\prime}. \label{eq:CRB}
\end{equation}

The 36 operators defined in Eq.~(\ref{eq:bilop}) satisfy the Lie algebra of the $U(6)$ group. There are also linear combinations of these operators that satisfy the Lie algebras of the $SU(3)$ and $SO(3)$ groups \cite{ibm}. This is referred to as the rotational limit of the IBM \cite{ibm,iach2}. Those linear combinations are given by the following angular momentum and quadrupole operators,
\begin{subequations}
\begin{eqnarray}
L_m & = & \sqrt{10} G_{m}^{(1)}(d,d), \quad m=-1,0,1, \\
Q_{m} & = & G_{m}^{(2)}(d,s) + G_{m}^{(2)}(s,d) - \dfrac{\sqrt{7}}{2} G_{m}^{(2)}(d,d), \quad k =-2,-1,\dots, 2. \label{su3_gen_LQ}
\end{eqnarray}
\end{subequations}

Using the commutation relations (\ref{eq:CRB}), it is straightforward to show that
\begin{subequations}
\begin{eqnarray}
[ L_0, L_1 ] & = & L_1, \\ \relax
[ L_0, L_{-1} ] & = & - L_{-1}, \\ \relax
[ L_1, L_{-1} ] & = & L_0, 
\end{eqnarray}
\end{subequations}
which correspond to the commutation relations for the spherical components of the angular momentum operator. Thus, defining the Cartesian components as
\begin{subequations}
\begin{eqnarray}
L^1 & = & \dfrac{\sqrt{2}}{2} \left( L_{-1} - L_1 \right), \\
L^2 & = & \dfrac{\sqrt{2}}{2} i \left( L_1 + L_{-1} \right), \\
L^3 & = & L_0, \label{eq:su2gen}
\end{eqnarray}
\end{subequations}
the Lie algebra of $SO(3)$, namely, $[L^i, L^j] = i \sum_k \epsilon^{ijk} L^k $, with $i,j,k = 1, 2, 3$, is recovered. Henceforth, the sum over repeated indices will be implicit, so expressions as $X^aY^a$ should be understood as $\sum_a X^aY^a$.

Following similar steps, the commutation relations of the Lie algebra of $SU(3)$ can be obtained. Starting from the definitions
\begin{eqnarray}
\label{eq:Ladder_su3}
\begin{array}{ll}
Y = \dfrac{2 \sqrt{2}}{3} Q_0, & T^3 = \dfrac12 L_0, \\[4mm]
T_+ = \dfrac{2}{\sqrt{3}} Q_2, & T_- = \dfrac{2}{\sqrt{3}}Q_{-2}, \\[4mm] 
U_+ = i \left[ \dfrac12 L_{-1} - \sqrt{\dfrac23} Q_{-1} \right], & U_- = i \left[ \dfrac12 L_{1} - \sqrt{\dfrac23} Q_{1} \right], \\[4mm]
V_+ = -i \left[ \dfrac12 L_{1} + \sqrt{\dfrac23} Q_{1} \right], & V_- = -i \left[ \dfrac12 L_{-1} + \sqrt{\dfrac23} Q_{-1} \right], 
\end{array}
\end{eqnarray}
it is quite easy to verify that
\begin{eqnarray}
[Y, T^3] & = & 0, \quad [Y, U_{\pm} ] = \pm U_{\pm}, \quad [Y, V_{\pm} ] = \pm V_{\pm}, \quad
[Y, T_{\pm} ] = 0, \nonumber \\ \relax
[T^3, U_{\pm} ] & = & \mp \dfrac{1}{2} U_{\pm}, \quad [T^3, V_{\pm} ] = \pm \dfrac{1}{2} V_{\pm}, \quad
[T^3, T_\pm] = \pm T_\pm, \nonumber \\ \relax
[T_+, T_-] & = & 2 T^3, \quad [T_+,V_+]= [T_+,U_-] = 0, \quad [T_+, V_-]=-U_-, \quad [T_+,U_+]=V_+, \nonumber \\ \relax
[T_-,V_-] & = & [T_-,U_+] = 0, \quad [T_-, V_+]= U_+, \quad [T_-,U_-]=-V_-, \nonumber \\ \relax
[U_+,U_-] & = & \dfrac{3}{2} Y - T^3, \quad [U_+,V_+]=0, \quad [U_+,V_-] = T_-, \nonumber \\ \relax
[U_-,V_-] & = & 0, \quad [U_-,V_+] = -T_+, \quad [V_+,V_-] = \dfrac{3}{2} Y + T^3.
\end{eqnarray}

Therefore, the Lie algebra of the $SU(3)$ group, $[T^a, T^b] = i f^{abc} T^c$, with $a,b,c = 1, \dots,8$, and
\begin{eqnarray}
\begin{array}{ll}
T^1 = \dfrac12 \left[ T_+ + T_- \right], \qquad &
T^2 = -\dfrac{i}{2} \left[ T_+ - T_- \right], \\[4mm]
T^4 = \dfrac12 \left[ U_+ + U_- \right], \qquad &
T^5 = \dfrac{-i}{2} \left[ U_+ - U_- \right], \\[4mm]
T^6 = \dfrac12 \left[ V_+ + V_- \right], \qquad &
T^7 = \dfrac{-i}{2} \left[ V_+ - V_- \right], \\[4mm]
T^8 = \dfrac{\sqrt{3}}{2} Y, &
\end{array}
\label{eq:su3gen}
\end{eqnarray}
is found \cite{greiner}.

Using the operators introduced in Eqs.~(\ref{eq:su2gen}) and (\ref{eq:su3gen}), the quadratic Casimir $C^{(2)}_{SU(3)}$ in the Hamiltonian (\ref{eq:h2su3}) reads
\begin{equation}
C^{(2)}_{SU(3)} = T^a T^a, \label{eq:su3casimir} 
\end{equation}
whose eigenvalue for the irrep $(\lambda, \mu)$ is \cite{greiner}
\begin{equation}
\dfrac13 \left[ \lambda^2 + \mu^2 + \lambda \mu + 3(\lambda + \mu) \right]. \label{eq:eigenv_SU3}
\end{equation}
The eigenvalues for the other Casimir operators in (\ref{eq:h2su3}) can be found in Refs.~\cite{ibm} and \cite{Pfeifer} and will not be given here.

In the $2$-body Hamiltonian with the dynamical symmetry of the chain algebra (\ref{eq:chain}), the energy difference between the rotational bands is given by the eigenvalue (\ref{eq:eigenv_SU3}). In order to gain more control on this difference, a $4$-body $SU(3)$-invariant term will be added to Hamiltonian (\ref{eq:h2su3}); this will be discussed in the following section.

\section{\label{sec:4body}Higher-order $SU(3)$-invariant terms in the IBM-1 Hamiltonian}

The $T^a$ operators listed in Eq.~(\ref{eq:su3gen}) are the generators of the Lie algebra of $SU(3)$ so they transform in the $8$-dimensional adjoint representation. The tensor product of two adjoint representations can be separated into the symmetric product $(8 \otimes 8)_S$ and the antisymmetric product $(8\otimes 8)_A$ as
\begin{subequations}
\label{eq:su3dec}
\begin{eqnarray}
(8\otimes 8)_S = 1 \oplus 8 \oplus 27, \\
(8\otimes 8)_A = 1 \oplus 10 \oplus \overline{10}.
\end{eqnarray}
\end{subequations}
where the different $SU(3)$ representations are denoted by their dimensions.

Prompted by the above decompositions, let
\begin{equation}
Q^{ab} = \alpha_1 Q_{(1)}^{ab} + \alpha_8 Q_{(8)}^{ab} + \alpha_{10+\overline{10}} Q_{(10 + \overline{10})}^{ab} + \alpha_{27} Q_{(27)}^{ab}, \label{eq:Q}
\end{equation}
be an $SU(3)$ $2$-body tensor operator constructed out of the product of two adjoints, where each term $Q_{(\mathrm{irrep})}^{ab}$ transforms according to the corresponding irrep indicated in the decomposition (\ref{eq:su3dec}) and $\alpha_{\mathrm{irrep}}$ are unknown coefficients.

Using the method to construct projection operators which can decompose any of the reducible finite-dimensional representation spaces of $SU(N)$ contained in the tensor product of two adjoint spaces into irreducible components introduced in Ref.~\cite{banda}, the $2$-body operators $Q_{(\mathrm{irrep})}^{ab}$ can straightforwardly written as
\begin{subequations}
\begin{equation}
Q_{(1)}^{ab} = \dfrac{1}{n^2-1} \delta^{ab} T^eT^e, \label{eq:Q1}
\end{equation}
\begin{equation}
Q_{(8)}^{ab} = \dfrac{n}{n^2-4} D^{abcd} T^cT^d + \dfrac{1}{n} F^{abcd} T^cT^d, \label{eq:Q8}
\end{equation}
\begin{equation}
Q_{(10 + \overline{10})}^{ab} = \dfrac12 (T^aT^b - T^bT^a) - \dfrac{1}{n} F^{abcd} T^c T^d, \label{eq:Q10}
\end{equation}
\begin{equation}
Q_{(27)}^{ab} = \dfrac{n+2}{4n}(T^aT^b + T^bT^a) - \dfrac{n+2}{2n(n+1)} \delta^{ab} T^eT^e - \dfrac{n+4}{4(n+2)} D^{abcd} T^cT^d + \dfrac{1}{4}(D^{acbd} + D^{adbc}) T^c T^d, \label{eq:Q27}
\end{equation}
\end{subequations}
where $n=3$ should be understood. For definiteness, $D^{abcd}$ and $F^{abcd}$ are given by
\begin{equation}
D^{abcd} = d^{abe}d^{cde}, \qquad F^{abcd} = f^{abe}f^{cde},
\end{equation}
where $d^{abc}$ is a symmetric rank-$3$ tensor.

Now, using the tensor operator $Q^{ab}$, Eq.~(\ref{eq:Q}), the $4$-body term $H^{(4)}$ in the IBM-1 Hamiltonian can readily be constructed as
\begin{equation}
H^{(4)} = Q^{ab} Q^{ab}, \label{eq:H4}
\end{equation}
so a Hamiltonian with up to $4$-body terms that respects the dynamical symmetry in (\ref{eq:chain}) is given by
\begin{equation}
H = c_1 + c_2 C^{(1)}_{U(6)} + c_3 C^{(2)}_{U(6)} + c_4 C^{(2)}_{SO(3)} - c_5 C^{(2)}_{SU(3)} - H^{(4)}, \label{eq:h4su3}
\end{equation}
where the signs of the last two summands have been conveniently chosen to describe the experimental spectra of the nucleus. The eigenfunctions of Hamiltonian (\ref{eq:h4su3}) have the same quantum numbers as the ones of Hamiltonian (\ref{eq:h2su3}), namely, $N$ for the boson number, $(\lambda,\mu)$ for a given $SU(3)$ irrep, $L$ for angular momentum, and $K$ for Elliott's quantum number. Notice that the first three coefficients $c_1,\ldots,c_3$ can be cast into a single one $E_0=c_1+c_2N+c_3N(N+1)$ since they contribute only to binding energies and not to excitation energies \cite{ibm}; the remaining coefficients depend on the physical system under consideration.

Before evaluating the tensor contraction indicated in $H^{(4)}$, an extra simplification can be exploited. Notice that 
\begin{eqnarray}
Q_{(10 + \overline{10})}^{ab} & = & \dfrac12 [T^a,T^b] - \dfrac{1}{n} F^{abcd} T^cT^d \nonumber \\
& = & \dfrac12 [T^a,T^b] - \dfrac{i}{2} f^{abe} T^e \nonumber \\
& = & 0,
\end{eqnarray}
where use of Eq.~(\ref{eq:Iden_FTT}) has been made.
 
Therefore, $Q^{ab}$ reduces to
\begin{equation}
Q^{ab} = \alpha_1 Q_{(1)}^{ab} + \alpha_8 Q_{(8)}^{ab} + \alpha_{27} Q_{(27)}^{ab}. \label{eq:qabf}
\end{equation}

Now, the tensor identities listed in Appendix A of Ref.~\cite{banda} and the commutation relation $[T^a,T^b]=if^{abc} T^c$ can be recursively used to rewrite $H^{(4)}$ in the form
\begin{equation}
H^{(4)} = \alpha_1^2 H_{1}^{(4)} + \alpha_8^2 H_{8}^{(4)} + \alpha_{27}^2 H_{27}^{(4)}, \label{eq:h4explicit}
\end{equation}
where
\begin{subequations}
\begin{eqnarray}
H_{1}^{(4)} & = & \dfrac{1}{64} \left( T^e T^e \right)^2, \label{eq:H1} \\
H_{8}^{(4)} & = & \dfrac{3}{5} D^{abcd} T^a T^b T^c T^d + \dfrac{1}{3} F^{abcd} T^a T^b T^c T^d, \label{eq:H8} \\
H_{27}^{(4)} & = & \dfrac{7}{8} \left( T^e T^e \right)^2 + i f^{abc} T^a T^b T^c + \dfrac{3}{4} T^e T^e - \dfrac{3}{5} D^{abcd} T^a T^b T^c T^d, \label{eq:H27}
\end{eqnarray}
\label{eq:Hs}
\end{subequations}
and $n=3$ has explicitly been set. Notice that not only pure $4$-body terms are contained in Eq.~(\ref{eq:Hs}), but also $2$- and $3$-body terms.

Equations (\ref{eq:Hs}), are they stand, can be algebraically worked out to readily obtain their eingenvalues by means of ordinary methods. Appendix \ref{sec:eigenvalues} describes one of such methods. However, a further simplification of relations (\ref{eq:Hs}) is achieved by using the definition of the quadratic Casimir operator of the Lie algebra of $SU(3)$, Eq.~(\ref{eq:su3casimir}), along with Eqs.~(\ref{eq:Iden_f3T}) and (\ref{eq:IdenD4T}). These terms become
\begin{subequations}
\label{eq:H_n}
\begin{eqnarray}
H_{1}^{(4)} & = & \dfrac{1}{64} \left[ C^{(2)}_{SU(3)} \right]^2, \label{eq:H1_n} \\
H_{8}^{(4)} & = & \dfrac{1}{5} \left[ C^{(2)}_{SU(3)} \right]^2 - \dfrac{3}{5} \, C^{(2)}_{SU(3)}, \label{eq:H8_n} \\
H_{27}^{(4)} & = & \dfrac{27}{40} \left[ C^{(2)}_{SU(3)} \right]^2 - \dfrac{9}{10} \, C^{(2)}_{SU(3)}. \label{eq:H27_n}
\end{eqnarray}
\end{subequations}

With the above expressions, the Hamiltonian (\ref{eq:h4su3}) can now be written as
\begin{eqnarray}
H & = & c_1 + c_2 C^{(1)}_{U(6)} + c_3 C^{(2)}_{U(6)} + c_4 C^{(2)}_{SO(3)} - c_5 C^{(2)}_{SU(3)} - \dfrac{\alpha_1^2}{64} \left[ C^{(2)}_{SU(3)} \right]^2 \nonumber \\
&  & \mbox{} - \dfrac{\alpha_8^2}{5} \left( \left[ C^{(2)}_{SU(3)} \right]^2 - 3 \, C^{(2)}_{SU(3)} \right) - \dfrac{\alpha_{27}^2}{10} \left( \dfrac{27}{4} \left[ C^{(2)}_{SU(3)} \right]^2 - 9 \, C^{(2)}_{SU(3)} \right). \label{eq:h4final}
\end{eqnarray}

\section{\label{sec:fit}Fitting experimental data for rotational bands: The one-parameter fit}

The energy gaps between rotational bands are described in Hamiltonian (\ref{eq:h4final}) by four parameters, namely, $c_5$, $\alpha_1$, $\alpha_8$, and $\alpha_{27}$. Important information about energy gaps can be learned by determining these parameters via a fit to data. For this task, two effective parameters can be defined, namely, one accompanying the quadratic Casimir and another one accompanying the quadratic Casimir squared, so these four parameters need be recombined accordingly. Unfortunately, the lack of experimental data does not allow one to perform an exploratory fit using the effective parameters, much less using all four parameters. A pragmatic approach is thus required in order to overcome this obstacle. The simplest approach, although perhaps not the optimal one, is to consider a one-parameter fit, either $c_5$, $\alpha_1^2$, $\alpha_8^2$, or $\alpha_{27}^2$.

The applicability of a one-parameter fit can be explored in two case examples: $^{156}$Gd and $^{234}$U. These nuclei are approximately described by the chain group (\ref{eq:chain}), provided some of their rotational bands can be treated as degenerate \cite{iach2}. Thus, in this approximation, and taking the spacing between non-degenerate rotational bands as the energy difference of the states with zero angular momentum, the $SU(3)$ term in the Hamiltonian (\ref{eq:h4final}) that yields the best fitting with one parameter for these spacings are
\begin{eqnarray}
\begin{array}{lll}
c_5 = \alpha_8 = \alpha_{27} = 0, & \quad \alpha_1^2 = 0.0593964 \, \mbox{keV} & \qquad \mbox{for \,} {}^{156}\mbox{Gd}, \\[4mm]
c_5 = \alpha_1 = \alpha_{27} = 0, & \quad \alpha_8^2 = 0.1500500 \, \mbox{keV} & \qquad \mbox{for \,} {}^{234}\mbox{U}.
\end{array}
\label{eq:Th_p}
\end{eqnarray}

\begin{figure}[ht]
\scalebox{0.35}{\includegraphics{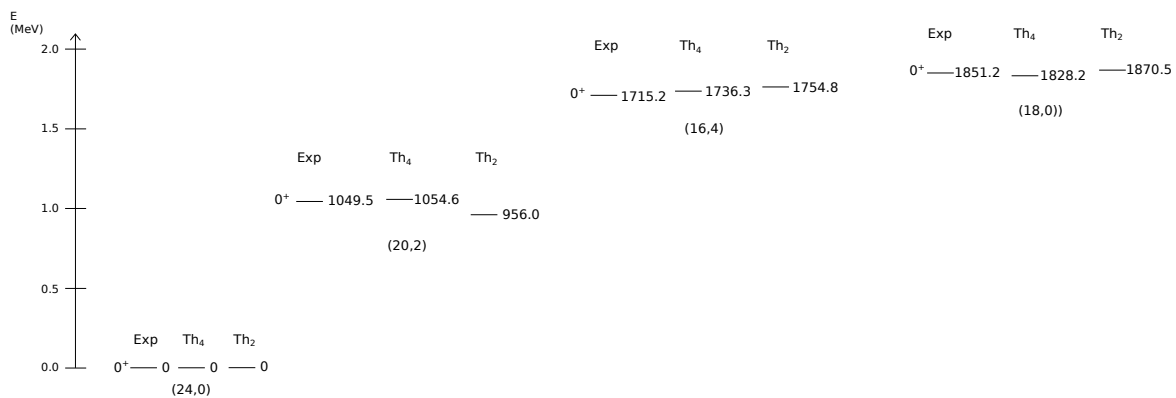}}
\caption{\label{fig:Gd_plot}Comparison between experimental and theoretical $SU(3)$ energy of the first excited states in $^{156}$Gd ($N=12$ in the IBM-1) of the rotation bands labelled by the indicated irreducible representation $(\lambda,\mu)$. Theoretical values of the parameters are given in Eq.~(\ref{eq:Th_p}), where $\mathrm{Th}_2$ and $\mathrm{Th}_4$ include up to $2$- and up to $4$-body contributions from Hamiltonians (\ref{eq:h2su3}) and (\ref{eq:h4final}), respectively. Experimental values are obtained from Ref.~\cite{ga}.}
\end{figure}

\begin{figure}[ht]
\scalebox{0.40}{\includegraphics{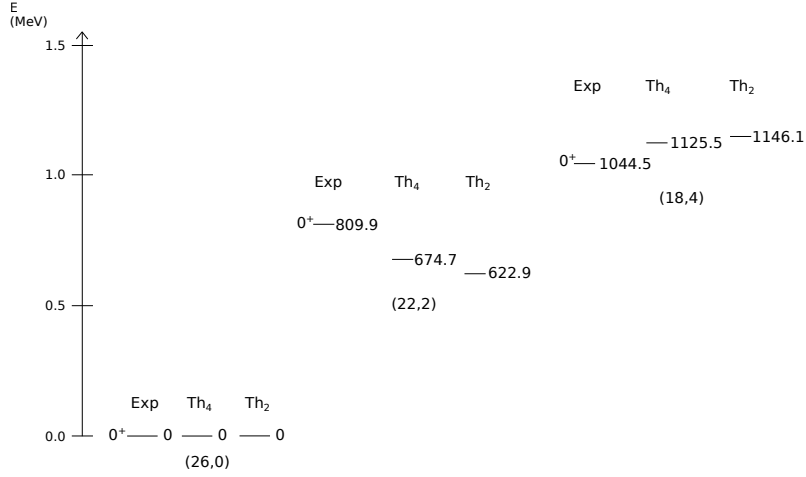}}
\caption{\label{fig:U_plot}Comparison between experimental and theoretical $SU(3)$ energy of the first excited states in $^{234}$U ($N=12$ in the IBM-1) of the rotation bands labelled by the indicated irreducible representation $(\lambda,\mu)$. Theoretical values of the parameters are given in Eq.~(\ref{eq:Th_p}), where $\mathrm{Th}_2$ and $\mathrm{Th}_4$ include up to $2$- and up to $4$-body contributions from Hamiltonians (\ref{eq:h2su3}) and (\ref{eq:h4final}), respectively. Experimental values are obtained from Ref.~\cite{u}.}
\end{figure}

In Figs.~\ref{fig:Gd_plot} and \ref{fig:U_plot} the comparison between the experimental energies of some excited states with zero angular momenta for $^{156}$Gd and $^{234}$U \cite{ga,u}, and the theoretical prediction given by the parameters in (\ref{eq:Th_p}) is made. In these figures, $\mathrm{Th_2}$ acccounts for effects of up to $2$-body terms from Hamiltonian (\ref{eq:h2su3}) whereas $\mathrm{Th}_4$ accounts for effects of up to $4$-body terms from Hamiltonian (\ref{eq:h4final}), with the best-fit parameters indicated in (\ref{eq:Th_p}). In the degenerate approximation these states correspond to the first excited states for each rotational band which are labelled by an irrep $(\lambda,\mu)$ of the $SU(3)$ group.

\section{\label{sec:concluding}Concluding remarks}

In this work, the $SU(3)$-invariant term $H^{(4)}$ with up to four-body operators, given in Eq.~(\ref{eq:H4}), is introduced in the standard Hamiltonian of the IBM-1, Eq.~(\ref{eq:h2su3}), for the chain symmetry $U(6) \supset SU(3) \supset SO(3)$. This term contributes to the energy spacing of the rotational bands that appear when nuclei are described by that specific symmetry. 

The energy term $H^{(4)}$ originates from the most general $2$-body operator $Q^{ab}$, which is constructed out of tensor product of two $SU(3)$ generators as given in Eq.~(\ref{eq:qabf}). The use of the projection operators introduced in Ref.~\cite{banda} applied to $Q^{ab}$ yields three main contributions to $H_4$, each one transforming according to well defined $SU(3)$ irreps. These contributions are $H_1, H_8$, and $H_{27}$ defined in Eqs.~(\ref{eq:Hs}).

After some simplifications, it turns out that $H_1, H_8$, and $H_{27}$ can be rewritten as linear combinations of the quadratic Casimir $C_{SU(3)}^{(2)}$ of the $SU(3)$ Lie algebra and its square as displayed in Eq.~(\ref{eq:H_n}). Thus, the Hamiltonian proposed in (\ref{eq:H4}), with up to fourth order $d$-boson interactions and that respects the dynamical symetry of the rotational limit of the IBM-1, becomes the Hamiltonian given in Eq.~(\ref{eq:h4final}). In that equation, the parameters $c_5$, $\alpha_1^2$, $\alpha_8^2$ and $\alpha_{27}^2$ describe the energy gaps of the rotational bands that are labeled according to distinct irrep of $SU(3)$. These four parameters can be recombined to obtain two effective parameters, one of them accompanying the quadratic Casimir $C_{SU(3)}^{(2)}$, and the other one accompanying the quadratic Casimir squared. Thus, the fitting to the experimental data can be performed using these two effective parameters. However, because of the lack of experimental data, a one-parameter fitting, either with $c_5$, $\alpha_1^2$, $\alpha_8^2$ or $\alpha_{27}^2$, was done. That fitting analysis is exemplified for the nuclei $^{156}$Gd and $^{234}$U. For the former the best fitting arise from $H_1$, choosing the specific values of the parameters as in (\ref{eq:Th_p}). It means that, for this nuclei, instead of the standard $2$-body IBM-1 Hamiltonian given in (\ref{eq:h2su3}) which is used in the literature, a better description is achieved by updating this same Hamiltonian to a $4$-body version using the Casimir $C_{SU(3)}^{(2)}$ squared instead. This situation where a higher-order interaction introduced into the IBM Hamiltonian improves the description of the experimental data is also reported in Refs.~\cite{Stefanescu,Heyde,Casten}. There it is shown how cubic terms in the IBM Hamiltonian can be used to describe features like triaxiality for some nuclei. 

Similarly, for $^{234}$U the best fitting occurs from $H_8$, with the parameters fixed as in (\ref{eq:Th_p}). Again, as in the $^{156}$Gd case, the $4$-body contribution represents a more accurate description of rotational bands than the standard $2$-body Hamiltonian when fitting to experimental data. 

A final closing remark about $H_1$, $H_8$, and $H_{27}$ is that, when they are used independently in the fitting analysis, the resulting energies are very close to each other, differing only in a few hundreds of eV. That subtlety comes from the different Casimir contribution $C_{SU(3)}^{(2)}$ in each term, as shown in (\ref{eq:Hs}). It means that the dominant part arises from the squared of $C_{SU(3)}^{(2)}$.

\begin{acknowledgments}
The authors are grateful to Consejo Nacional de Ciencia y Tecnolog{\'\i}a (Mexico) for partial support.
\end{acknowledgments}

\appendix

\section{\label{sec:iden}Some useful identities of $SU(3)$ generators}

In this section, some useful identities involving $SU(3)$ generators are obtained to help simplify relations (\ref{eq:Hs}).

Let $T^a$ be the generators of the Lie algebra of the $SU(n)$ group, with commutation relations $[T^a,T^b]=if^{abc}T^c$.

A simple algebraic manipulation yields 
\begin{eqnarray}
\dfrac{1}{n} F^{abcd} T^c T^d & = & \dfrac{1}{2n} f^{abe} f^{cd e} \left( \{ T^c, T^d \} + [T^c, T^d] \right) \nonumber \\
& = & \dfrac{i}{2n} f^{abe} f^{cd e} f^{cdg} T^g. \label{eq:F2T}
\end{eqnarray}

Using the identity \cite{djm95},
\begin{equation}
f^{abc} \, f^{abd} = n \, \delta^{cd}, \label{eq:spinflavor_1}
\end{equation}
allows one to rewrite Eq.~(\ref{eq:F2T}) as
\begin{equation}
\dfrac{1}{n} F^{abcd} T^c T^d = \dfrac{i}{2} f^{abe} T^e. \label{eq:Iden_FTT}
\end{equation}

In the same vein as the previous case, it can be shown that
\begin{equation}
i f^{abc} T^a T^b T^c = - \dfrac{n}{2} T^e T^e, \label{eq:Iden_f3T}
\end{equation}
which yields a simplified version of the second summand of Eq.~(\ref{eq:H27}).

Now, reducing the terms containing four $T$'s is a little bit more involved. For this task, the projection operators $\pr{3}$ and $\pr{4}$ for $SU(n)$ introduced in Ref.~\cite{banda}, namely,
\begin{eqnarray}
\left[ \pr{3} \right]^{c_1c_2b_1b_2} & = & \dfrac{n+2}{4n} \left( \delta^{c_2 b_1} \delta^{c_1b_2} + \delta^{c_1b_1} \delta^{c_2b_2} \right)
+ \dfrac{1}{4} \left( D^{c_1b_1c_2b_2} + D^{c_2b_1c_1b_2} \right) \nonumber \\
&  & \mbox{} - \dfrac{n+4}{4(n+2)} D^{c_1 c_2 b_1 b_2} - \dfrac{n+2}{2n(n+1)} \delta^{c_1c_2} \delta^{b_1b_2},
\end{eqnarray}
and
\begin{eqnarray}
\left[ \pr{4} \right]^{c_1c_2b_1b_2} & = & \dfrac{n-2}{4n} \left( \delta^{c_2b_1} \delta^{c_1b_2} + \delta^{c_1b_1} \delta^{c_2b_2} \right)
- \dfrac{1}{4} \left( D^{c_1b_1c_2b_2} + D^{c_2b_1c_1b_2} \right) \nonumber \\
&  & \mbox{} + \dfrac{n-4}{4(n-2)} D^{c_1c_2b_1b_2} + \dfrac{n-2}{2n(n-1)} \delta^{c_1c_2} \delta^{b_1b_2},
\end{eqnarray}
are useful. First, notice that
\begin{equation}
\left[ \pr{3} \right]^{c_1c_2b_1b_2} + \left[ \pr{4} \right]^{c_1c_2b_1b_2} = \dfrac{1}{2} \left( \delta^{c_2b_1} \delta^{c_1b_2} + \delta^{c_1b_1} \delta^{c_2b_2} \right) - \dfrac{n}{n^2-4} D^{c_1c_2b_1b_2} - \dfrac{1}{n^2-1} \delta^{c_1c_2} \delta^{b_1b_2}. \label{eq:sum_proj}
\end{equation}

For $n=3$, the $SU(3)$ case, the projector $\left[ \pr{4} \right]^{c_1c_2b_1b_2}$ vanishes. Thus, from Eq.~(\ref{eq:sum_proj}), it follows that
\begin{equation}
D^{c_1b_1c_2b_2} + D^{c_2b_1c_1b_2} + D^{c_1c_2b_1b_2} = \dfrac{1}{3} \left( \delta^{c_2b_1} \delta^{c_1b_2} + \delta^{c_1b_1} \delta^{c_2b_2} + \delta^{c_1c_2} \delta^{b_1b_2} \right), \label{eq:DSU3_iden}
\end{equation}
where the tensor $D^{c_1c_2b_1b_2}$ is defined in the Lie algebra of the $SU(3)$ group.

Using the commutator $ [T^a,T^b]=if^{abc}T^c$ and the identity \cite{djm95}
\begin{equation}
d^{abc} d^{ade} f^{bdf} = \dfrac{n^2-4}{2n} f^{cef}, \label{eq:spinflavor_2}
\end{equation}
the tensor contraction indicated in Eq.~(\ref{eq:H8}) reduces to
\begin{equation}
D^{abcd} T^aT^bT^cT^d = \dfrac{1}{3} (T^e T^e)^2 + \dfrac{1}{4} T^e T^e, \label{eq:IdenD4T} 
\end{equation}
which ultimately can be expressed in powers of the quadratic Casimir of $SU(3)$.

\section{\label{sec:eigenvalues}A general method to compute the eigenvalues of $H^{(4)}$}

The analyses of nuclei with a large number of valence nucleons involve $SU(3)$ irreps of high dimensionality. The numerical approach to obtain the corresponding eigenvalues becomes a rather difficult computational task. In this section a general procedure to find the eigenvalues $\lambda_k$ of the operator $H^{(4)}$ defined in Eq.~(\ref{eq:h4explicit}) is outlined. Recalling that each contribution to $H^{(4)}$ transforms according to a specific $SU(3)$ representation simplifies the analysis.

First, the eigenvalues $\lambda_1$ of $H_{1}^{(4)}$ for an irrep $(\lambda,\mu)$ are the simplest to obtain. Since $T^e T^e$ is the quadratic Casimir of the Lie algebra of $SU(3)$, from Eq.~(\ref{eq:eigenv_SU3}), $\lambda_1$ simply reads
\begin{equation}
\lambda_1 = \dfrac{1}{192} \left[ \lambda^2 + \mu^2 + \lambda\mu + 3(\lambda+\mu) \right]^2.
\end{equation}

Now, the eigenvalues $\lambda_8$ and $\lambda_{27}$ of $H_{8}^{(4)}$ and $H_{27}^{(4)}$ given in Eqs.~(\ref{eq:H8}) and (\ref{eq:H27}), respectively, can be obtained by computing the eigenvalues of the explicit matrix expressions for these operators in a given irrep $(\lambda,\mu)$. A few of those eigenvalues, mainly for the lowest irreps, have been computed with the help of the algorithm described in Ref.~\cite{algorithm_su3} and are listed in Table \ref{tab:H_eigen}. However, due to the huge computational effort required in the numerical evaluations, it is more convenient to have analytical expressions instead that can help complete Table \ref{tab:H_eigen}. For this purpose, a linear regression analysis using the values formerly obtained through the algorithm of Ref.~\cite{algorithm_su3} can be performed. The conventional technique of a linear regression (see for example Ref.~\cite{bishop}) requires a linear combination of $M$ functions $f_{k}(\lambda,\mu)$, with $f_0(\lambda,\mu)=1$, to fit the data set $\mathcal{D} = \{(\lambda_{r}^{(1)},(\lambda_1,\mu_1)), \dots,(\lambda_{r}^{(N)},(\lambda_N,\mu_N))\}$, with $r=8,27$. Here $\lambda_{r}^{(n)}$ indicates the eigenvalue for the corresponding representation $(\lambda_n,\mu_n)$.

Let $\bm{\lambda_r}$ be the vector
\begin{eqnarray}
\bm{\lambda_r} = \begin{pmatrix}
\lambda_{r}^{(1)} \\
\vdots \\
\lambda_{r}^{(N)}
\end{pmatrix},
\end{eqnarray}
in such a way that the function 
\begin{equation}
g_r(\lambda,\mu) = \bm{m}_r \cdot \bm{f}(\lambda,\mu), \label{predictive}
\end{equation}
minimizes the sum of squared errors given by 
\begin{equation}
\sum_{n=1}^N \left[ \lambda_{r}^{(n)} - g_r(\lambda_n,\mu_n) \right]^2.
\end{equation}

Here,
\begin{eqnarray}
\bm{m}_r = \left( \Phi^T \Phi \right)^{-1} \Phi^T \bm{\lambda_r}, \qquad \qquad
\bm{f}(\lambda,\mu) = \begin{pmatrix}
f_{0}(\lambda,\mu) \\
\vdots \\
f_{M-1}(\lambda,\mu)
\end{pmatrix},
\end{eqnarray}
and $\Phi$ represents the design matrix and $\Phi^T$ its transpose. Specifically, $\Phi$ reads \cite{bishop},
\begin{eqnarray}
\Phi = \begin{pmatrix}
f_0(\lambda_1,\mu_1) & \dots & f_{M-1}(\lambda_1,\mu_1) \\
\vdots &  & \vdots \\
f_0(\lambda_N,\mu_N) & \dots & f_{M-1}(\lambda_N,\mu_N)
\end{pmatrix}.
\end{eqnarray}

Since the entries of Table \ref{tab:H_eigen} are generated by an unknown function of $\lambda,\mu$, the right choice of the set of functions $f_k(\lambda,\mu)$ can perfectly fit these data without overfitting. 

The choice of
\begin{equation}
\bm{f}(\lambda,\mu) = ( 1, \mu, \dots, \mu^4, \lambda, \lambda \mu, \dots, \lambda \mu^4,
\dots, \lambda^4, \lambda^4 \mu, \dots, \lambda^4 \mu^4 )^T,
\end{equation}
for the regression analysis yields
\begin{subequations}
\begin{eqnarray}
\lambda_8 & = & g_8(\lambda,\mu) = \dfrac{1}{45} \left[ \lambda^2 + \lambda (3 + \mu) + \mu (3 + \mu) - 9 \right] \left[ \lambda^2 + \lambda (3 + \mu) + \mu (3 + \mu) \right], \label{eq:H8_eigen} \\
\lambda_{27} & = & g_{27}(\lambda,\mu) = \dfrac{3}{40} \left[ \lambda^2 + \lambda (3 + \mu) + \mu (3 + \mu) \right] \left[ \lambda^2 + \lambda (3 + \mu) + (-1 + \mu) (4 + \mu) \right]. \label{eq:H27_eigen}
\end{eqnarray}
\end{subequations}
The above expressions exactly reproduce the numerical values obtained with the help of the algorithm of Ref.~\cite{algorithm_su3} and are also useful to complete Table \ref{tab:H_eigen}. These entries also agree in full with the eigenvalues obtained from Eqs.~(\ref{eq:H8_n}) and (\ref{eq:H27_n}), so the overall analysis is consistent.

\begingroup
\squeezetable
\begin{table}
\caption{\label{tab:H_eigen}Eigenvalues $\lambda_8$ and $\lambda_{27}$ of the operators $H_{8}^{(4)}$ and $H_{27}^{(4)}$ defined in Eqs.~(\ref{eq:H8}) and (\ref{eq:H27}), respectively, for the corresponding $SU(3)$ irrep $(\lambda,\mu)$.}
\begin{center}
\begin{tabular}{ccccc}
\hline\hline
$SU(3)$ irrep $(\lambda,\mu)$ & \qquad & $\lambda_8$ & \qquad & $\lambda_{27}$ \\ \hline
$(0,1)$ & & $-4/9$ & & $0$ \\
$(0,2)$ & & $2/9$ & & $9/2$ \\
$(0,3)$ & & $18/5$ & & $189/10$ \\
$(0,4)$ & & $532/45$ & & $252/5$ \\
$(0,5)$ & & $248/9$ & & $108$ \\
$(0,6)$ & & $54$ & & $405/2$ \\
$(1,1)$ & & $0$ & & $27/8$ \\
$(1,2)$ & & $112/45$ & & $72/5$ \\
$(1,3)$ & & $80/9$ & & $315/8$ \\
$(1,4)$ & & $108/5$ & & $432/5$ \\
$(1,5)$ & & $392/9$ & & $1323/8$ \\
$(1,6)$ & & $704/9$ & & $288$ \\
$(2,2)$ & & $8$ & & $36$ \\
$(2,3)$ & & $170/9$ & & $153/2$ \\
$(2,4)$ & & $1702/45$ & & $1449/10$ \\
$(2,5)$ & & $68$ & & $252$ \\
$(2,6)$ & & $5092/45$ & & $2052/5$ \\
$(3,3)$ & & $36$ & & $1107/8$ \\
$(3,4)$ & & $2842/45$ & & $2349/10$ \\
$(3,5)$ & & $4672/45$ & & $15111/40$ \\
$(3,6)$ & & $162$ & & $1161/2$ \\
$(4,4)$ & & $504/5$ & & $1836/5$ \\
\hline\hline
\end{tabular}
\end{center}
\end{table}
\endgroup

\end{document}